\documentclass[aps,preprintnumbers,nofootinbib]{revtex4}
\usepackage{epsfig,graphics}
\usepackage{amsfonts,amssymb,amsmath}            
\usepackage{pifont}                              
\usepackage{amsthm}
\usepackage{subfigure}
\def\identity{\leavevmode\hbox{\small1\kern-3.8pt\normalsize1}}

\newcommand{\ket}[1]{\left |  #1 \right\rangle}
\newcommand{\bra}[1]{\left \langle #1  \right |}

\newcommand{\Tr}{\rm tr}
\newcommand{\ketbra}[2]{|#1\rangle\!\langle#2|}
\theoremstyle{plain}

\newcounter{pp}

\begin{document}

\title{Variable Bias Coin Tossing}
\date{\today}

\author{Roger \surname{Colbeck}}
\email[]{r.a.colbeck@damtp.cam.ac.uk}
\affiliation{Centre for Quantum Computation,
             DAMTP,
             Centre for Mathematical Sciences,
             University of Cambridge,
             Wilberforce Road,
             Cambridge CB3 0WA, U.K.}

\author{Adrian \surname{Kent}}
\email[]{a.p.a.kent@damtp.cam.ac.uk}
\affiliation{Centre for Quantum Computation,
             DAMTP,
             Centre for Mathematical Sciences,
             University of Cambridge,
             Wilberforce Road,
             Cambridge CB3 0WA, U.K.}

\begin{abstract}
Alice is a charismatic quantum cryptographer who believes her 
parties are unmissable; Bob is a\footnote{relatively} 
glamorous string theorist who believes he is an indispensable guest.    
To prevent possibly traumatic collisions of self-perception and reality, their
social code requires that decisions about invitation or acceptance be
made via a cryptographically secure {\it variable bias coin toss} (VBCT).
This generates a shared random bit by the toss of a coin whose bias is
secretly chosen, within a stipulated range, by one of the parties; the
other party learns only the random bit.  
Thus one party can secretly influence the outcome, while both
can save face by blaming any negative decisions on bad luck. 

We describe here some cryptographic VBCT protocols whose security is
guaranteed by quantum theory and the impossibility of superluminal
signalling, setting our results in the context of a general discussion
of secure two-party computation. We also briefly discuss other
cryptographic applications of VBCT.
\end{abstract}

\maketitle

\section{Introduction}

\subsection{Background} 

The discoveries of quantum cryptography \cite{Wiesner} and
provably secure quantum key distribution
\cite{BB84,Ekert,Mayers2,LoChau2,ShorPreskill}  motivated a general
search for protocols which implement interesting cryptographic tasks
in a way that can be guaranteed secure by quantum theory (for example \cite{Ambainis,Ambainis&,ArrighiSalvail,BCJL,CGS,Lo,Mochon2,Spekkens&Rudolph,Spekkens&Rudolph_CSWCF,Aharonov&2,HardyKent}),
by the impossibility of superluminal signalling \cite{Kent_relBCshort,
Kent_relBC, BHK},  or both.   The full cryptographic power of these
physical principles is presently unknown: ideally, one would like to
generate either a provably secure protocol or a no-go theorem for
every interesting task.

There are at least three significant types of cryptographic security 
which apply to protocols based on physics:  
\begin{enumerate}
\item {\it Unconditional} security, where the impossibility of
  useful cheating (i.e.\ learning private information or influencing
  the outcome of the protocol beyond what is permitted by an honest
  input) is guaranteed by the laws of physics.
\item {\it Cheat-evident} security, where at least one party can
  usefully cheat,  but the laws of physics guarantee that any cheating
  will eventually be detected with certainty.
\item {\it Cheat-sensitive} security
  \cite{Aharonov&2,HardyKent,Spekkens&Rudolph_CSWCF}, where  at least
  one party can usefully cheat, but the laws of physics guarantee that
  any such cheating will be detected with non-zero probability.  
\end{enumerate}
In this paper, we focus mainly on unconditional
security, but also consider an interesting cheat-evident protocol. 

We follow the standard convention that a protocol is secure 
provided that it protects honest parties from cheats.  Thus, for 
the two-party protocols considered here, we do not require that a 
protocol offers any protection if both parties cheat.  Instead, we
simply guarantee to each party that if they follow the protocol as
prescribed, they will be protected.
To be more precise, the parties are guaranteed protection against
{\it useful} cheating.  It is not necessary in mistrustful 
cryptography to prevent every possible kind of deviation from
a protocol.  What is required is some form of guarantee that any deviations
which go undetected give no advantage to the party who deviates:
i.e.\ that the deviating party gains no unauthorized information about
the other party's inputs and no illegitimate influence over the protocol's outcome.  
For example, in a relativistic coin tossing protocol in which the parties
are supposed to independently supply random bits $a$ and $b$ and the 
coin toss outcome is $c = a \oplus b$, there is no way to guarantee 
to $A$ that $B$'s bit $b$ was genuinely randomly chosen (or vice versa).  However,
this does not matter: as long as at least one party is honest, the outcome $c$ 
is random.  Thus, though an honest party has no guarantee that they will
detect all deviations from the protocol by the other party, they do have
a guarantee that, if the protocol produces a coin toss outcome, it will
be fair.   

Most work on quantum cryptography to date has considered non-relativistic
protocols, in which the parties' locations are completely unconstrained
and their communications may effectively be assumed instantaneous.  
However, for at least two important tasks, strong coin tossing and
bit commitment, we know that protocols which rely on the impossibility
of superluminal signalling are more powerful than their non-relativistic
counterparts.   Strong coin tossing --- in which two
mistrustful parties want to create a shared random bit whose randomness
is guaranteed --- is trivial to implement using relativistic
signalling constraints (see e.g.\ Ref.\ \cite{Kent_CTBC}), but cannot
be securely implemented using non-relativistic protocols
\cite{Ambainis&,Kitaev}.
Non-relativistic quantum bit commitment has also been shown to 
be impossible \cite{Mayers,LoChau,KMP}.  On the other hand, 
there exist relativistic protocols \cite{Kent_relBC,Kent_relBCshort}
which are (unlike any non-relativistic protocol) provably secure
against classical attacks \cite{Kent_relBC}, and also provably immune to Mayers-Lo-Chau
attacks \cite{Kent_relBC}: it is conjectured that they are also secure against general
quantum attacks.

Relativistic protocols require each party to be able to 
send and receive communications from at least two separated
locations.
The separation between an individual party's
communication devices must be considerably greater than the 
separation between their device and the other party's 
nearest device.
For instance, if Alice uses locations $A_1$ and $A_2$
and Bob uses locations $B_1$ and $B_2$, the distance
$d( A_1 , A_2 )$ must be considerably greater than
$d(A_1 , B_1 )$ and $d( A_2 , B_2 )$. 

Using quantum communications and storing and manipulating quantum states
in order to implement a cryptographic protocol is clearly an 
inconvenience: quantum technology seems likely to be more costly
and less robust than its classical counterpart for the foreseeable
future.   The constraints imposed by relativistic 
protocols may also in some circumstances be a significant
inconvenience.  For example, if 
two parties occupy small secure sites which are widely
separated, and trust nothing outside their secure sites,
they cannot run a relativistic protocol securely without relocating
or extending their sites.  
Of course, in both cases, the compensating advantage is a
guarantee of unconditional security which cannot be obtained
by other means.  
It is also worth stressing that relativistic protocols do not
require either party to trust that the other is located where
they claim to be: each party can guarantee security by knowing
their own locations and by recording the times at which they 
send and receive signals.   Nor does relativistic cryptography
necessarily require large-scale separation: in principle, two
parties could implement a relativistic protocol by placing
two credit card sized secure devices next to one another.   

\subsection{Secure computation} 

The main focus of this paper is to consider protocols for the task
of variable bias coin tossing (VBCT) between two parties.  
Roughly speaking --- we give precise definitions below ---
a secure VBCT protocol generates a shared
random bit as though by a biased coin, whose bias is secretly
chosen by one of the parties to take some value within a prescribed
range.  This is the simplest case of the more general task of
carrying out a variable bias $n$-faced die roll, in which one
of $n$ possible outcomes is randomly generated as though by a 
biased die, whose bias (i.e.\ list of outcome probabilities)
is secretly chosen by one of the parties to 
take some value within a prescribed convex set.
Variable bias coin tossing and die rolling are themselves special
cases of secure two-party computations.  To understand their 
significance, it is helpful to locate them within a general
classification of secure computation tasks.  

A general secure classical computation involves $N$ parties, 
labelled by $i$ in the range $1 \leq i \leq N$, who each
have some input, $x_i$, and wish to compute some (possibly
non-deterministic) functions of their inputs, with the $i$-th party
receiving as output $f_i(x_1 , \ldots , x_N )$.  We call this a classical computation
because the inputs and outputs are classical, although we allow such 
computations to be implemented by protocols which involve the processing of quantum states.  
All of the computations we consider in this paper are classical in this sense (although
most of the protocols we discuss involve quantum information processing), 
so we will henceforth refer simply to computations, with the term ``classical''
taken as understood.   A perfectly secure
computation guarantees, for each $i$, each subset $J \subseteq \{ 1 , \ldots , N \}$,
and each set of possible inputs $x_i$ and $\{x_j\}_{j \in J}$, 
that if the parties $J$ do indeed input $\{x_j\}_{j \in J}$ and then collaborate, 
they can gain no more information about the input $x_i$ than 
what is implied by $\{x_j\}_{j \in J}$
and $\{f_j (x_1 , \ldots , x_N )\}_{j \in J}$. 

We restrict attention here to two types of two-party computation: {\it
  two-sided} computations in which the outputs prescribed for each
party are identical, and {\it one-sided} computations in which one
party gets no output.  We use the term {\it single function
  computations} to cover both of these types, since in both cases only
one function need be evaluated.  We can classify single function
computations by the number of inputs, by whether they are
deterministic or random, and by whether one or two parties receive the
output.

We are interested in protocols whose unconditional security is 
guaranteed by the laws of physics.  In particular, as is standard
in these discussions, we do not allow any security arguments
based on technological or computational bounds: each party 
allows for the possibility that the other may have 
arbitrarily good technology and arbitrarily powerful quantum computers. 
Nor do we allow any reliance on mutually trusted
third parties or devices.   We also make the standard 
assumptions that Alice and Bob are the only participants in the
protocol --- i.e.\ that there is no interference by third parties ---
and that they have noiseless communication channels. 

The known results for secure computations are summarized
below.

{\bf Zero-input computations:} \qquad Secure protocols 
for zero-input deterministic computations or zero-input random
one-sided computations can be trivially constructed, since the
relevant computations can be carried out by one or both parties separately.  
The most general type of zero-input two-sided random secure computation 
is a biased $n$-faced secure die roll.  This can be implemented 
with unconditional security by generalizing the well-known relativistic 
protocol for a secure coin toss (see e.g.\ Ref.\ \cite{Kent_CTBC}).  

\smallskip

{\bf One-input computations:}  \qquad Secure protocols for deterministic one-input computations are
trivial; the party making the input can always choose it to generate
any desired output on the other side, so might as well compute the
function on their own and send the output directly to the other party.  

The non-deterministic case is of interest.  For one-sided computations,
where the output goes to the party that did not make the input, the
most general function is a one-sided variable bias $n$-faced die roll.
The input simply defines a probability distribution over the outputs.
In essence, one party chooses one from a collection of biased
$n$-faced dice to roll (the members of the collection are known to
both parties).  The output of the roll goes to one party
only, who has no other information about which die was chosen.  

It is known that some computations of this type are impossible.  A
special case of these computations defines a version of oblivious
transfer (OT), in which Alice inputs a bit, Bob inputs nothing, Bob
receives Alice's bit with probability half, and otherwise receives the
outcome {\it fail}.  Rudolph \cite{Rudolph} has shown that no
non-relativistic quantum protocol can securely implement this task,
and his argument trivially generalizes to the relativistic case.

The two-sided case of a non-deterministic one-input function we call a
variable bias $n$-faced die roll.  This --- and particularly the
two-faced case, a variable bias coin toss ---  is the subject of the
present paper.   We will give a protocol that implements the task with
unconditional security for a limited range of biases, another which
implements any range of biases, but achieves only cheat-evident
security, and two further protocols that allow any range of biases and
which we conjecture are unconditionally secure.

\smallskip

{\bf Two-input computations:} \qquad Lo \cite{Lo} considered the task
of finding a secure non-relativistic quantum protocol for a two-input,
deterministic, one-sided function. He showed that if the protocol
allows Alice to input $i$, Bob to input $j$, and Bob to receive
$f(i,j)$, while giving Alice no information on $j$, then Bob can also
obtain $f(i,j')$ for all $j'$.  For any cryptographically nontrivial
computation, there must be at least one $i$ for which knowing
$f(i,j')$ for all $j'$ gives Bob more information than knowing
$f(i,j)$ for just one value of $j$.  As this violates the definition
of security for a secure classical computation, it is impossible to
implement any cryptographically nontrivial computation securely. 
Lo's proof as stated applies to non-relativistic protocols, and
extends trivially to relativistic protocols. 
We hence conclude that all secure two-input deterministic one-sided
cryptographically nontrivial computations are impossible.

Lo also noted that some secure two-input deterministic, two-sided
non-relativistic quantum computations are impossible, because they
imply the ability to do non-trivial secure two-input, deterministic
one-sided computations.  This  argument also extends trivially to
relativistic protocols. 

As far as we are aware, neither existence nor no-go results are
presently known for secure two-input non-deterministic computations.

\smallskip

Table \ref{fns} summarizes these results.

\begin{table}
\begin{tabular}{|l|l|c|c|l|}
\hline
No input&Deterministic &$ \checkmark $  &Trivial\\
        &Random       one-sided         &$ \checkmark $  &Trivial\\
        &Random       two-sided         &\checkmark   &Biased $n$-faced die roll\\
\hline
One-input &Deterministic &$\checkmark $  &Trivial\\
        &Random       one-sided         &(\ding{55})&One-sided
variable bias $n$-faced die roll\\
        &Random       two-sided         &$\checkmark^*$ &Variable bias
$n$-faced die roll\\
\hline
Two-input &Deterministic one-sided         &\ding{55}    &c.f.\ Lo\\
        &Deterministic two-sided        &(\ding{55})  &c.f.\ Lo\\
        &Random       one-sided        &?            &\\
        &Random       two-sided        &?            &\\
\hline
\end{tabular}
\caption{Functions computable securely in two-party computations using
  (potentially) both quantum and relativistic protocols.  \checkmark
  indicates that all functions of this type are possible, \ding{55}
  indicates that all functions of this type are impossible,
  $\checkmark^*$ indicates that the conjectures made later in this paper
  imply that all functions of this type are possible, (\ding{55}) 
  indicates that some functions of
  this type are impossible, and ? indicates no known result.}
\label{fns}
\end{table}

\section{Variable bias coin tossing} \label{secvbct}

\subsection{Introduction}

We now specialize to the task of variable bias coin tossing (VBCT), the
simplest case of a one-input, random, two-sided computation.
We seek protocols whose security is guaranteed based on the laws of physics.  
We distinguish {\it relativistic} protocols, which
rely on the impossibility of superluminal signalling, 
from {\it non-relativistic} protocols, which do not. 
We also distinguish {\it quantum} protocols, which require
quantum information to be generated and exchanged, from {\it
  classical} protocols, which can be implemented using classical
information alone.

The aim of a VBCT protocol is to provide
two mistrustful parties with the outcome of a biased coin toss.
We label the possible outcomes by $0$ and $1$ and define the
{\it bias}
to be the probability $p_0$ of outcome $0$.  
The protocol should allow one party, by convention Bob, to fix the bias to take
any value within a pre-agreed range, $p_{\rm min} \leq p_0 \leq p_{\rm max}$.
Roughly speaking --- modulo epsilonics and technicalities which we discuss below ---
the protocol should guarantee to both parties that the biased coin toss outcome is 
genuinely random, in the sense that Bob's only way of influencing the outcome
probabilities is through choosing the bias, while Alice has no way of influencing
the outcome probabilities at all.    It should also guarantee to Bob that 
Alice can obtain no information about his bias choice beyond what she can 
infer from the coin toss outcome alone.

To illustrate the uses of VBCT, consider a situation in which 
Bob may or may not wish to accept Alice's invitation to a party, in a future world 
in which social protocol decrees that his decision\footnote{Naturally,
a similar protocol, in which Alice chooses the bias, governs the decision
about whether or not an invitation is issued.}     
is determined by a variable bias coin toss in which he chooses the bias within a
prescribed range, let us say $p_{\rm min} =
\frac{1}{11} \leq p_0 \leq p_{\rm max} = \frac{10}{11}$.
Alice, who is both self-confident and a Bayesian, believes prior to 
the coin toss that the probability of Bob not wishing to accept
is $10^{-n}$, for some fairly large value of $n$.   If Bob does indeed wish
to accept, he can choose $p_0 = \frac{10}{11}$, ensuring a
high probability of acceptance. 
If he does not, he can choose $p_0 = \frac{1}{11}$, ensuring a 
low probability of acceptance.  
If the invitation is declined, this social protocol allows both parties
to express regret, ascribing the outcome to bad luck rather than to
Bob's wishes.  Alice's posterior probability estimate of Bob's not
wishing to attend is approximately $ 10^{-n+1}$, i.e.\ still close to
zero.

For another illustration of the uses of VBCT, suppose 
that Bob has a large secret
binary dataset of size $N$.  For example, this might be
a binary encoding of a high resolution satellite image.  
He is willing to sell Alice a noisy image of the dataset
with a specified level of random noise.   Alice is willing to purchase if 
there is some way of guaranteeing, at least to within tolerable
bounds, that the noise is at the specified level and that it was genuinely
randomly generated.  In particular, she would like some guarantee that 
constrains Bob so that he cannot selectively choose the noise so
as to obscure a significantly sized component of the 
dataset which he (but not necessarily she) knows to be especially 
interesting.   Let us suppose also that the
full dataset will eventually become public, so that Alice will be able
to check the noisy image against it, and that she will be able to 
enforce suitably large penalties against Bob if the 
noisy and true versions turn out not to be appropriately related. 
They may proceed by agreeing parameters $p_{\rm min}$ 
and $p_{\rm max} = 1 - p_{\rm min}$, and then running
a variable bias coin toss for each bit in the image, with
Bob choosing $p_0 = p_{\rm min}$ if the bit is $1$ and 
$p_0 = p_{\rm max}$ if the bit is $0$.   Following this 
protocol honestly provides Alice with the required randomly
generated noisy image.  On the other hand, if Bob deviates
significantly from these choices for more than $O( \sqrt{N} )$
of the bits, Alice will almost certainly be able to 
unmask his cheating once she acquires the full dataset.  

\subsection{Definitions}

A VBCT protocol is defined by a prescribed series of classical
or quantum communications between two parties, Alice and Bob.  
If the protocol is relativistic, it may also require that the parties
each occupy two or more appropriately located sites, and may stipulate
which sites each communication should be made from and to.  
The protocol's definition includes bias parameters
$p_{\rm min}$ and $p_{\rm max}$, with $p_{\rm min} < p_{\rm max}$, and
may also include one or more security parameters $N_1 , \ldots ,
N_r$.  It accepts a one bit input from one party, Bob, and 
must result in both parties receiving the same output, one of the
three possibilities $0$, $1$ or ``abort''.   The output ``abort'' can
arise only if at least one of the parties refuses to complete the
protocol honestly.  

We follow the convention that Bob can fix $p$ to be $p_{\rm min}$ or
$p_{\rm max}$ by choosing inputs $1$ or $0$ respectively (so that an
input of bit value $b$ maximizes the probability of
output $b$).  He can thus fix $p$ anywhere in the range $p_{\rm min}
\leq p \leq p_{\rm max}$ by choosing the input randomly with an
appropriate weighting.  Since any VBCT protocol gives Bob this
freedom, we do not require a perfectly secure protocol to exclude
other strategies which have the same result: i.e.\ a perfectly secure
protocol may allow any strategy of Bob's which causes $p_0$ to lie in
the given range, so long as no other security condition is 
violated.\footnote{Similar statements hold, with appropriate 
epsilonics, for secure protocols: see below.}  This motivates
the following security definitions.  

We say the protocol is {\it secure} if the following conditions
hold when at least one party honestly follows the protocol.
Let $p_0$ be the probability of the output being $0$, and 
$p_1$ be the probability of the output being $1$. 
Then, regardless of the strategy that a dishonest party may  
follow during the protocol, we have $ p_0  \leq p + \epsilon (N_1 ,
\ldots , N_r )$ and $ p_1 \leq  (1-p) + \epsilon (N_1 , \ldots , N_r
)$, where $p_{\rm min} \leq p \leq p_{\rm max}$ and the protocol
allows Bob to determine $p$ to take any value in this range.   
Alice has probability less than $\zeta(N_1,\ldots,N_r)$ of obtaining more 
than $I + \delta(N_1 , \ldots , N_r)$ bits of
information about the value of $p$ determined by Bob's input, where
$I$ is the information implied by the coin toss outcome.  In addition,
if Bob honestly follows the protocol and legitimately aborts before
the coin toss outcome is known\footnote{We
  take this to be the point at which both
  parties have enough information (possibly distributed between their
  remote agents) to determine the outcome.}, then Alice has
probability less than $\zeta(N_1 , \ldots , N_r)$ of obtaining more
than $\delta(N_1 , \ldots , N_r )$ bits of information about Bob's input. 

(We should comment here on a technical detail that will be relevant
to some of the protocols we later consider.   It turns out, in some 
of our protocols, to be possible and useful for Bob to make 
supplementary security tests even after both parties have received
information which would determine the coin toss outcome.   The protocols
are secure whether or not these supplementary tests are made, in 
the sense that the security criteria hold as the security parameters
tend to infinity.  However, the supplementary tests increase the 
level of security for any fixed finite value of the security parameters.

We need slightly modified definitions to cover this case, since the
output of the protocol is defined to be ``abort'' if Bob aborts
after carrying out supplementary security tests.   
If Bob honestly follows a protocol with supplementary tests,
and legitimately aborts after the coin toss outcome is
determined, then we require
that Alice should have probability less than $\zeta(N_1 , \ldots ,
N_r)$ of obtaining more than $\delta (N_1 , \ldots , N_r )$ extra bits
of information --- i.e.\ beyond what is implied by the coin toss
outcome --- about Bob's input.

Note that introducing supplementary security tests may allow Alice to 
follow the protocol honestly until she obtains the coin toss 
outcome, and then deliberately fail the supplementary tests in order to
cause the protocol to abort.  However, this gives her no useful extra
scope for cheating.  In any type of VBCT protocol, she can always 
follow the protocol honestly and then refuse to abide by the outcome: 
for example, she can decide not to invite Bob to her party, even if 
the variable bias coin toss suggests that she should.  This unavoidable
possibility has the same effect as her causing the
protocol to abort after the coin toss outcome is determined.)

In all the above cases, we require $\delta (N_1 , \ldots , N_r ) \rightarrow 0$, 
$\epsilon (N_1 , \ldots , N_r ) \rightarrow 0$ and
$\zeta(N_1,\ldots,N_r)\rightarrow 0$ as the $N_i \rightarrow \infty$.
We say the protocol is {\it perfectly secure} for some fixed values $N_1 ,
\ldots , N_r$ if the above conditions hold with $\epsilon(N_1 , \ldots
, N_r ) = \delta (N_1 , \ldots , N_r ) = \zeta(N_1 , \ldots , N_r ) = 0$.

Suppose now that one party is honest and the other party fixes 
their strategy (which may
be probabilistic and may depend on data received during the
protocol) before the protocol commences, and suppose that the 
probability of the protocol aborting, given this strategy,
is less than $\epsilon'$.  
Since the only possible outcomes are $0$, $1$ and ``abort'', 
it follows from the above conditions that, if Bob inputs $1$, we have  
$p_{\rm min} - \epsilon (N_1 , \ldots , N_r ) - \epsilon' 
< p_0 \leq p_{\rm min} + \epsilon (N_1 , \ldots , N_r )$
and 
$(1 - p_{\rm min}) - \epsilon (N_1 , \ldots , N_r ) - \epsilon'
< p_1 \leq (1 - p_{\rm min}) + \epsilon (N_1 , \ldots , N_r )$.  
Similarly, if Bob inputs $0$, we have  
$p_{\rm max} - \epsilon (N_1 , \ldots , N_r ) - \epsilon' 
< p_0 \leq p_{\rm max} + \epsilon (N_1 , \ldots , N_r )$
and 
$(1 - p_{\rm max}) - \epsilon (N_1 , \ldots , N_r ) - \epsilon' 
< p_1 \leq (1 - p_{\rm max}) + \epsilon (N_1 , \ldots , N_r )$.
In other words, unless a dishonest party is willing to 
accept a significant risk of the protocol aborting, they 
cannot cause the outcome probabilities for $0$ or $1$ to 
be significantly outside the allowed range.  Moreover, no
aborting strategy can increase the probability of $0$ or $1$
beyond the allowed maximum. 

For an {\it unconditionally secure} VBCT protocol, the above
conditions hold assuming only that the laws of physics are correct. In
a {\it cheat-evidently secure} protocol, if any of the above
conditions fail, then the non-cheating party is guaranteed to detect
this, again assuming only the validity of the laws of physics.

\section{Some cryptographic background}

The VBCT protocols we discuss below require both parties
to set up separated sites from which they can send and
receive communications, and rely on the impossibility 
of sending signals faster than light between these sites.
Most of them also require quantum information to be 
transmitted and manipulated.  In other words, the
protocols are (in most cases) both quantum and relativistic.  

Some of the protocols we consider use bit commitment
as a subprotocol.  Specifically, they use the 
relativistic bit commitment protocol RBC2 described in
Ref.\ \cite{Kent_relBC}. This protocol has been
proven secure against all classical attacks.  It has also been
proven immune to the Mayers-Lo-Chau attack which renders
non-relativistic quantum bit commitment protocols insecure
\cite{Kent_relBC}.  It is conjectured to be secure against general
quantum attacks.

For completeness, we include here brief reviews of the simplest
scenario for relativistic cryptography and of the notion
of bit commitment, as previously set
out in Ref.\ \cite{Kent_relBC}. 

\subsection{Review of relativistic cryptography} 

We assume that physics takes place in flat Minkowski spacetime, with
the Minkowski causal structure.  This is not exactly correct, of
course --- general relativity and experiment tell us that spacetime is
curved --- but it is true to a good enough approximation for any
protocol implemented on or near Earth.  In principle, our protocol's
timing constraints should take into account the error in the
approximation.  Other than this, the known corrections to the 
local causal structure arising from general relativity do 
not affect our security analyses.  

Like all physics-based cryptographic protocols, the security of
the relativistic protocols we consider ultimately relies 
on the (approximate) validity of the underlying 
physical model.  In principle, they would be vulnerable
to a malicious adversary with the power
to distort spacetime significantly (yet surreptitiously!) 
in the region of the protocol.  For example, a protocol using
separated sites in London and Sydney would be vulnerable if
one party reconfigured the local spacetime geometry so that
the two cities became geodesically separated by, say, 
$\approx 10^{-3}$ light sec instead of $\approx 4 \times 10^{-2}$ light sec.   
For obvious reasons, we do not take this possibility --- or
other scenarios involving exotic and speculative general relativistic 
phenomena \cite{Kent_relBCshort} --- seriously at present.  

We use units in which the speed of light is unity and choose
inertial coordinates, so that the minimum possible time
for a light signal to go from one point in space to another
is equal to their spatial separation.  
We consider a cryptographic scenario in which coordinates
are agreed by Alice and Bob, who also agree on two points\footnote{
This discussion generalizes in an obvious way to 
cover protocols, such as the protocol VBCT1 considered below,
which require Alice and Bob to control three or more 
separate sites.}
$\underline{x}_1$ and $\underline{x}_2$.  Alice and Bob are 
required to erect laboratories, including sending and receiving
stations, within an agreed distance $\delta$ of the points, where 
$ D = | \underline{x}_1 - \underline{x}_2 | \gg \delta $.  
These laboratories need not be restricted in
size or shape, except that they must not overlap. 

We refer to the laboratories in the vicinity of $\underline{x}_i$ as 
$A_i$ and $B_i$, for $i = 1$ or $2$. To avoid unnecessarily
proliferating notation, we use the same labels
for the agents (sentient or otherwise) assumed to 
be occupying these laboratories.  
The agents $A_1$ and $A_2$ 
may be separate individuals or devices, but we assume 
that they are collaborating with complete mutual trust and with
completely prearranged agreements on how to proceed, to the extent 
that for cryptanalytic purposes
we can identify them together simply as a single entity,
Alice ($A$); similarly $B_1$ and $B_2$ are identified as Bob ($B$).  

Note that, in many situations where any sort of 
cryptography (not necessarily quantum or relativistic) is 
employed, this sort of identification is in any case 
natural and indeed necessary.  Governments, companies and other
organizations are represented by many agents at distributed
locations.  When two such organizations, $A$ and $B$, exchange data via
a cryptographic protocol, each organization typically has to 
assume that several of its own agents are trustworthy.  The aim of 
the protocol is then to ensure that, provided this assumption this
correct, neither organization obtains unauthorized information
from the other.   

It is perhaps also worth stressing that requiring $A$ and $B$
to trust their own agents or devices is entirely different from introducing
a third party trusted by both $A$ and $B$.  
While the first assumption (which we make) is natural and often necessary, 
the second (which, to reiterate, we do not allow) would be illegitimate in the context of 
the present discussion.  (Many mistrustful cryptographic tasks, including
all those we consider here, can be trivially implemented if $A$ and $B$
can both rely on the same trusted third party.) 

As usual in defining a cryptographic scenario for a protocol
between mistrustful parties, we suppose
Alice and Bob each trust absolutely the security and integrity of their own
laboratories, in the sense that they are confident that 
all their sending, receiving and analysing devices 
function properly and also that nothing within their laboratories
can be observed by outsiders.  They also have confidence in the
locations of their own laboratories in the agreed coordinate
system, and in clocks set up within their laboratories.    
However, neither of them trusts any third party or channel or device
outside their own laboratory.  

To ensure in advance that their clocks are synchronized and that
their communication channels transmit at sufficiently near light
speed, the parties may check that test signals sent out from each of
Bob's laboratories receive a response within time $4 \delta$ from
Alice's neighbouring laboratory, and vice versa.  However, the parties
need not disclose the precise locations of their laboratories in order
to implement the protocol.  Nor need Alice or Bob take it on trust
that the other has set up laboratories in the stipulated regions. 
(A protocol which required such trust would, of course, be fatally
flawed.)  Each party can verify that the other is not significantly
deviating from the protocol by checking the times at which signals
from the other party arrive.  These arrival times, together with the
times of their own transmissions, can be used to guarantee that
particular specified pairs of signals, going from Alice to Bob and
from Bob to Alice, were generated independently. This guarantee is
all that is required for security.

Given a laboratory configuration as above, one can set out
precise timing constraints for all communications in a protocol in
order to ensure the independence of all pairs of signals which are
required to be generated independently.  We may use the time coordinate in
the agreed frame to order the signals in the protocol.  (Without such a
convention there would be some ambiguity, since the time ordering is
frame dependent).  

We also assume that $A_1$ and $A_2$ either have, or can securely
generate, an indefinite string of random bits. This string
is independently generated and identically distributed, with
probability distribution defined by the protocol, and is denoted ${\bf
  x}\equiv\{x_i\}$.  Similarly, $B_1$ and $B_2$ share a random string
${\bf y}\equiv\{y_i\}$.  These random strings will be used to make all
random choices as required by the protocol: as $A_1$ and $A_2$, for
instance, both possess the same string, ${\bf x}$, they know the
outcome of any random choices made during the protocol by the other. 
We also assume the existence of secure authenticated pairwise channels between
the $A_i$ and between the $B_i$.  We do not assume that these channels are necessarily
unjammable: we need only stipulate that, if an honest party fails to receive the signals as
required during any of the protocols we discuss, they abort the protocol.

\subsection{Brief review of bit commitment}

Roughly speaking --- precise definitions can be found in,
for example, Ref.\ \cite{Kent_relBC}  
--- bit commitment is the cryptographic version
of a safe and key.  In the commitment phase
of a bit commitment protocol, Alice supplies Bob with 
data that commit her to the value
of a bit, without allowing Bob to infer that value.  This corresponds
to locking the bit in the safe and handing it to Bob.
In the unveiling phase, which takes place some time after commitment, if and
when Alice wishes, she supplies Bob with further data (the key in our
analogy) in order to reveal the value of the bit to which she
committed.

Recently, it was shown \cite{Kent_relBC} 
that there exist relativistic bit commitment protocols which are provably  
unconditionally secure against classical attack, in the 
sense that the laws of classical physics (including special
relativity) imply that neither party can cheat, 
regardless of the technology or computing power available to them. 
Mayers, Lo and Chau had earlier shown \cite{Mayers,LoChau}
that non-relativistic quantum bit commitment protocols are 
necessarily insecure, by constructing an explicit attack that
allows the committer to cheat against any protocol which is
secure against the receiver.   However, it was shown in 
Ref.\ \cite{Kent_relBC} that the relativistic bit commitment
protocols described there are immune to Mayers-Lo-Chau type
attacks.  It is conjectured that they are in fact
unconditionally secure, i.e.\ that they are immune to
general quantum attacks.  

\section{VBCT protocols}

\subsection{Protocol VBCT1}

We consider first a simple relativistic quantum protocol, which
implements VBCT with unconditional security, for a limited range of biases.
The protocol requires each party to have agents located at
three appropriately separated sites.  

\begin{enumerate}

\item \label{poisson} $B_1$, $B_2$ and $B_3$ agree a random number $n$ chosen from a
  Poisson distribution with large mean (or other suitable distribution).  

\item $A_1$ sends a sequence of qubits $\{\ket{\phi_i}\}$ to $B_1$, where
each $\ket{\phi_i} \in \{ \ket{ \psi_0},\ket{\psi_1} \}$ is chosen
independently with probability half each, using the
random string ${\bf x}$. The states $\ket{ \psi_0}$ and $\ket{\psi_1}$
are agreed between Alice and Bob prior to the
protocol, and the qubits are sent at regular intervals according to a
previously agreed schedule, so that all the agents involved can
coordinate their transmissions.     

\item $B_1$ receives each qubit and stores it.

\item \label{seq} $A_2$ tells $B_2$ the sequence of states $\{ \ket{\phi_i} \}$
sent, choosing the timings so that $A_1$'s quantum communication of the qubit $\ket{\phi_i}$
is spacelike separated from $A_2$'s classical communication of its identity.  
$B_2$ relays these communications to $B_1$.

\item \label{Bcheck1} On receipt, $B_1$ measures his stored states to
  check that they are correctly described by $A_2$.  If any error
  occurs he aborts.

\item \label{announce} 
$B_3$ announces to $A_3$ that the $n$th state will be used for the coin
toss.  This announcement is made at a point spacelike separated from
the $n$th rounds of communication between $A_1$ and $B_1$ and $A_2$ 
and $B_2$.  $A_3$ reports the value of $n$ to $A_1$ and $A_2$.

\item \label{Bmeasure} 
$B_1$ performs the measurement on $\ket{\phi_n}$ that optimally
distinguishes $\ket{\psi_0}$ from $\ket{\psi_1}$, and then
reveals to $A_1$ that this is the state that will be used, along with
a bit $b$.  If his measurement is indicative of the state being
$\ket{\psi_{b'}}$, then Bob should select $b=b'$ if he wants outcome
0, or else select $b=\bar{b'}$.  Let Alice's random choice for the
$n$th state be $\ket{\psi_a}$: recall that $A_2$ reported the value
of $a$ to $B_2$ in step \ref{seq}.

\item Some time later, $A_1$ receives from $A_3$ the value of $n$ 
sent by $B_3$, confirming that $B_1$ was committed to guess the $n$th state, 
and $B_1$ receives from $B_2$ the value of $a$ sent by $A_2$.
The outcome of the coin toss is $c= a \oplus b$.
\end{enumerate}

It will be seen that this protocol is a variant of 
the familiar relativistic protocol for ordinary coin tossing.
As in that protocol, Alice and Bob simultaneously
exchange random bits.  However, Alice's bit is here encoded in
non-orthogonal qubits, which means that Bob can obtain
some information about its value. 
Bob uses this information to affect the bias of the coin toss.  

We use the bit $w$ to represent Bob's wishes, with $w=0$
representing Bob trying to produce the outcome $0$ by guessing
correctly, and $w=1$ representing him trying to produce the
outcome $1$ by guessing wrongly. Security requires that  
\begin{align}
\label{criterion}
p(w|a,b,c) \approx p(w|c) \, .  
\end{align}
Perfect security requires equality in the above equation.

\subsubsection{Bob's strategy}

The choice of $n$ need not be fixed by Bob at the start of the
protocol: for example, it could be decided during the protocol by
using an entangled state shared by the $B_i$.  However, we may assume $B_3$
sends a classical choice of $n$ to $A_3$ ($A_3$ will measure any
quantum state he sends immediately in the computational basis, and hence
we may assume, for the purposes of security analysis, that $B_3$ carries 
out this measurement).  $B_3$'s announcement
of $n$ is causally disconnected from the sending of the $n$th state to
$B_1$ and of its identity to $B_2$.  Therefore, no matter how it is
selected, it does not depend on the value of the $n$th state.  
While it could be generated in such a way as to depend on
some information about the sequence of states previously received, 
these states are uncorrelated with the $n$th state if Alice follows
the protocol.  Such a strategy thus confers no advantage, and we
may assume, for the purposes of security analysis, that the 
the choice of $n$ is generated by an algorithm independent of
the previous sequence of states.     
We may also assume that $n$ is generated in such a way that $B_1$ and $B_2$ can obtain
the value of $n$ announced by $B_3$ with certainty: if not, their task
is only made harder. 
In summary, for the purposes of security analysis, we may assume that
$B_3$ announces a classical value of $n$, pre-agreed with $B_1$ and
$B_2$ at the beginning of the
protocol.  

$B_1$ is then committed to making a guess of the value of the 
$n$th state: if he fails to do so then Alice knows Bob has cheated. 
$B_1$'s best strategy is thus to perform some measurement on the $n$th
state and use the outcome to make his guess.  We define
$\ket{\psi_0}=\cos{\frac{\theta}{2}}\ket{0}+\sin{\frac{\theta}{2}}\ket{1}$
and
$\ket{\psi_1}=\cos{\frac{\theta}{2}}\ket{0}-\sin{\frac{\theta}{2}}\ket{1}$,
where $0\leq\theta\leq\frac{\pi}{2}$. 
Let the projections defining the optimal measurement be $P_0$ and $P_1$. 
We say that the outcome corresponding to $P_0$ is `outcome 0', and similarly
for the outcome corresponding to $P_1$.  Without loss of generality,
we can take outcome 0 to correspond to the most likely state Alice
sent being $\ket{\psi_0}$ and similarly outcome 1 to correspond to
$\ket{\psi_1}$. Bob's probability of guessing correctly is then given
by,
\begin{align}
p_B = \frac{1}{2}\left(\bra{\psi_0}P_0\ket{\psi_0}+\bra{\psi_1}P_1\ket{\psi_1}\right) \, .
\end{align}
This is maximized for $P_0$ and $P_1$ corresponding to measurements in
the $\ket{\pm}$ basis, where
$\ket{\pm}=\frac{1}{\sqrt{2}}(\ket{0}\pm\ket{1})$.  The maximum value
is,
\begin{align}
\label{p_Bmax}
p_B^{max}=\frac{1}{2}\left( 1+\sin{\theta}\right).
\end{align}

It is easy to see that the security criterion (\ref{criterion}) is always 
satisfied.
The minimum probability of Bob guessing correctly is always
$1-p_B^{max}$, which he can attain by following the same strategy but
associating outcome $b'$ with a guess of $\bar{b'}$.   
The possible range of biases are those between
$p_{\rm min}=\frac{1}{2} \left( 1+\sin{\theta} \right)$ and
$p_{\rm max}=\frac{1}{2} \left( 1-\sin{\theta} \right)$.
The protocol thus implements VBCT for all values of $p_{\rm min}$
and $p_{\rm max}$ with $p_{\rm min} + p_{\rm max} =1$ (and no others).  

\subsubsection{Security against Alice}

Security against Alice is ensured by the fact that
$B_1$ tests $A_2$'s statements about the identity of the states
sent to $B_1$.   

We seek to show that if Alice attempts to alter the probability of $B_1$
measuring $0$ or $1$ with his measurement in step \ref{Bmeasure}, then in the
limit of large $n$, either the probability of her being detected tends
to $1$, or her probability of successfully altering the probability
tends to zero.  Note that it may be useful for Alice to alter the
probabilities in either direction: if she increases the
probability that $B_1$ guesses correctly, she learns more information
about Bob's bias than she should; if she
decreases it, she limits Bob's ability to affect the bias.

We need to show that if on the $i$-th round, $B_1$ receives
state $\rho_i$, for which the probability of outcome $0$ differs from those
dictated by the protocol, then the probability of $B_1$ not detecting
Alice cheating on this state is strictly less than $1$.

$B_1$'s projections are onto $\{\ket{+},\ket{-}\}$ for the
$n$th state.
Alice's cheating strategy must ensure that for some subset of the states
she sends to $B_1$, there is a different probability of his
measurement giving outcome $0$.  Suppose that $\rho_i$ satisfies
\begin{align}
\label{overlap}
\bra{+}\rho_i\ket{+}&=p_{\rm max}+\delta_1\\
&=p_{\rm min}+\delta_2\, ,
\end{align}
where $\delta_1,\delta_2\neq 0$.
Then, if $B_1$ was to instead test Alice's honesty, the state
which maximizes the probability of Alice passing the test, among those
satisfying (\ref{overlap}), is 
\begin{align}
(p_{\rm max}+\delta_1)^{\frac{1}{2}}\ket{+}+(1-p_{\rm max}-\delta_1)^{\frac{1}{2}}\ket{-},
\end{align}
and she should declare this state to be whichever of $\ket{\phi_0}$ or
$\ket{\phi_1}$ maximizes the probability of passing Bob's test.  We have
\begin{align}
\left(( p_{\rm max} ( p_{\rm max} + \delta_1 ))^{\frac{1}{2}} + 
( ( 1-p_{\rm max}) (1-p_{\rm max} - \delta_1 ))^{\frac{1}{2}}\right)^2 \leq 
1-\delta_1^2\, ,
\end{align}
and a similar equation with $p_{\rm min}$ replacing $p_{\rm max}$ and
$\delta_2$ replacing $\delta_1$.
Hence the probability of passing Bob's test is at most $1 - \delta^2$, 
where $\delta = \min (|\delta_1| , |\delta_2| )$.  
In order to affect $B_1$'s measurement probabilities with significant
chance of success, there must be a significant fraction of states
satisfying (\ref{overlap}).  If a fraction $\gamma$ of states satisfy
(\ref{overlap}) with $\min (|\delta_1| , |\delta_2| ) \geq \delta$ for
some fixed $\delta >0$, then this cheating strategy
succeeds with probability at most $\gamma(1-\delta^2)^{\gamma n}$.
Hence, for any $\delta$, $\gamma$, the probability of this technique
being successful for Alice can be made arbitrarily close to $0$ if Bob
chooses the mean of the Poisson distribution used in step
\ref{poisson} (and hence the expected value of $n$) to be sufficiently
large.

Note that, as this argument applies state by state to the $\rho_i$, 
it covers every possible strategy of Alice's: in particular, the
argument holds whether or not the sequence of qubits she sends
is entangled.   

We hence conclude that the protocol is asymptotically secure against Alice.

\subsection{Protocol VBCT2}

We now present a relativistic quantum VBCT protocol which allows any range of 
biases, but achieves only cheat-evident security rather than unconditional
security.

\begin{enumerate}
\item \label{first step} $B_1$ creates $N$ states, each being either 
  $\ket{\psi_0}=\alpha_0\ket{00}+\beta_0\ket{11}$ or
  $\ket{\psi_1}=\alpha_1\ket{00}+\beta_1\ket{11}$, with
  $\{\alpha_0,\alpha_1,\beta_0,\beta_1\}\in\mathbb{R}^+$, 
  $\alpha_0^2 > \alpha_1^2$, and
  $\alpha_i^2+\beta_i^2=1$.  The states are chosen with probability
  half each.  In the unlikely event that all the states are the same,
  $B_1$ rejects this batch and starts again.   $B_1$ uses the shared
  random string ${\bf y}$ to make his random choices, so that $B_1$
  and $B_2$ both know the identity of the $i$-th state.  $B_1$ sends
  the second qubit of each state to $A_1$.  The values of
  $\alpha_0,\beta_0,\alpha_1$ and $\beta_1$ are known to both Alice
  and Bob.   We define the bias of the state $\ket{\psi_i}$ to be
  $\alpha_i^2$, and write $p_{\rm min} = \alpha_1^2$ and  $p_{\rm max}
  = \alpha_0^2$.
\item \label{Alice decides} Alice decides whether to test Bob's honesty
  ($z=1$), or to trust him ($z=0$).  She selects $z=0$ with probability
  $2^{-M}$. $A_1$ and $A_2$ simultaneously inform $B_1$ and $B_2$ of
  $z$, $A_2$'s communication being spacelike separated from the
  creation of the states by $B_1$ in step \ref{first step}.
\item \begin{enumerate} \item If $z=1$, $B_1$ sends all
  of his qubits and their identities to $A_1$, while $B_2$ sends the
  identities to $A_2$.  $A_1$ can then verify that they are as claimed and
  if so, the protocol returns to step \ref{first step}.  If not, she aborts
  the protocol.
  \item \label{Bcheck} If $z=0$, $B_1$ randomly chooses a state to use
  for the coin toss from amongst those with the bias he wants. 
$B_2$ simultaneously informs $A_2$ of
  $B_1$'s choice. \end{enumerate}
\item $A_1$ and $B_1$ measure their halves
  of the chosen state in the $\{\ket{0},\ket{1}\}$ basis, and this
  defines the outcome of the coin toss.
\renewcommand{\labelenumi}{(\arabic{enumi}.}
\item \label{extratest}
  As an optional supplementary post coin toss security test, $B_1$ may 
  ask $A_1$ to send all her remaining qubits back to him, except for her half
  of the state selected for the coin toss. He can then perform projective
  measurements on these states to check that they correspond to those
  originally sent.)

\end{enumerate}

An intuitive argument for security of this protocol is as follows.
On the one hand, as $M\rightarrow\infty$, the protocol is secure
against Bob since, in this limit, he always has to convince Alice that
he supplied the right states which he can only do if he has been honest.  
But also, in the limit $N\rightarrow\infty$, we expect the protocol to
be secure against Alice, since in this limit, she cannot gain any more
information about the bias Bob selected than can be gained by
performing the honest measurement.  

The protocol can only provide cheat-evident security rather than
unconditional security, since there are useful cheating strategies open to
Alice, albeit ones which will certainly be detected.  One such strategy is
for $A_1$ to claim that $z=0$ on some state, while $A_2$ claims that
$z=1$.  This allows Alice to determine Bob's desired bias, since $B_1$
will tell $A_1$ the state to use, and $B_2$ will tell $A_2$ its
identity.  However, this cheating attack will be exposed once $B_1$
and $B_2$ communicate.  

(Technically, Alice has another possible attack:
she can follow the protocol honestly until she learns the outcome, and
then intentionally try to fail Bob's tests in step \ref{extratest} by
altering her halves of the remaining states in some way.  By so doing,
she can cause the protocol to abort after the coin toss outcome is
determined.  However, as discussed in Section \ref{secvbct}, this
gives her no advantage.)

\subsubsection{Security against Alice}

Assume Bob does not deviate from the protocol.  $A_2$ must announce
the value of $z$ without any information about the current batch of
states sent to $A_1$ by $B_1$.  Alice
therefore cannot affect the bias: once a given batch is
accepted, she cannot affect $B_1$'s measurement
probabilities on any state he chooses for the coin toss.
While Alice's choices of $z$ need not be classical bits determined before
the protocol and shared by the $A_i$, we may assume, for the purposes
of security analysis, that they are, by the same argument used 
in analysing Bob's choice of $n$ in VBCT1.     

Once Bob has announced the state he wishes to use for
the coin toss, though, Alice can perform any measurement
on the states in her possession in order to gain information about
Bob's chosen bias.   It would be sufficient to show that any such
attack that provides significant information is almost certain to be
detected by Bob's tests in step \ref{Bcheck}; if so, the existence of
such attacks would not compromise the cheat-evident security of the
protocol.  In fact, a stronger result holds: Alice cannot gain
significant information by such attacks.  From her perspective, if Bob
selects a $\ket{\psi_0}$ state for the coin toss, the (un-normalized)
mixed state of the remaining $(N-1)$ qubits is,
\begin{align}
\tilde{\sigma}_0 \equiv \sum_{m=0}^{N-2} \, \sum_{{i_1 , \ldots , i_{N-1} \in \{ 0, 1\}} \atop 
{\sum_{j=1}^{N-1} i_j = (N-1-m)}  }
 \rho_{i_1} \otimes  \rho_{i_2} \otimes \cdots \rho_{i_{N-1}}   \, , 
\end{align}
while if Bob selects a $\ket{\psi_1}$ state for the
coin toss, the (un-normalized) mixed state of the remaining $(N-1)$ qubits is 
\begin{align}
\tilde{\sigma}_1 \equiv \sum_{m=1}^{N-1} \, \sum_{{i_1 , \ldots , i_{N-1} \in \{ 0, 1\}} \atop 
{\sum_{j=1}^{N-1} i_j = (N-1-m)}  }
 \rho_{i_1} \otimes  \rho_{i_2} \otimes \cdots \rho_{i_{N-1}}   \, , 
\end{align}
where
\begin{align*}
\rho_i = {\Tr}_B ( \ketbra{\psi_i}{\psi_i} ) \qquad {\rm for~}i=0,1
\, .
\end{align*}
We will use $\sigma_0$ and $\sigma_1$ to denote the normalized
versions of $\tilde{\sigma}_0$ and $\tilde{\sigma}_1$ respectively.

We have
\begin{align}
D(\rho_0\otimes\sigma_0,\rho_1\otimes\sigma_1)&\leq
D(\rho_0\otimes\sigma_0,\rho_1\otimes\sigma_0)+
D(\rho_1\otimes\sigma_0,\rho_1\otimes\sigma_1)\\
\end{align}
where $D(\rho,\sigma) = \frac{1}{2}\Tr | \rho - \sigma |$ is the trace distance
between $\rho$ and $\sigma$. 
As $N \rightarrow \infty$, we have $D(\sigma_0 , \sigma_1) \rightarrow 0$ 
and so $D(\rho_0\otimes\sigma_0,\rho_1\otimes\sigma_1)
\rightarrow D(\rho_0 , \rho_1)$. 
Since the maximum probability of distinguishing two states is a
function only of the trace distance \cite{Helstrom}, 
the maximum probability of distinguishing $\rho_0\otimes\sigma_0$ from 
$\rho_1\otimes\sigma_1$ tends, as $N \rightarrow \infty$,
to the maximum probability of distinguishing $\rho_0$ from $\rho_1$.
The measurement that attains this maximum is that dictated by the
protocol.  We hence conclude that, in the limit of large $N$, the excess
information Alice can gain on Bob's chosen bias by using
any cheating strategy tends to zero.

\subsubsection{Security against Bob}

We now consider Bob's cheating possibilities, assuming
that Alice does not deviate from the protocol.    
To cheat, Bob must achieve a bias outside the
range permitted.  Let us suppose he wants to ensure that 
the outcome probability of $0$ satisfies $p_0 \geq p_{\rm max} + \delta$,
for some $\delta > 0$ (the case $p_1 \geq 1-p_{\rm min} + \delta$ can
be treated similarly), and let us suppose this can be achieved
with probability $\delta' > 0$.  

For this to be the case, there must be some cheating strategy
(possibly including measurements) which, with probability $\delta'$,
allows $B_2$ to identify a choice of $i$ from the relevant batch of
$N$ qubits such that the state $\rho_i$ of $A_1$'s $i$-th qubit then
satisfies
\begin{align}
\label{zer}
\bra{0} \rho_i \ket{0} \geq p_{\rm max} + \delta .
\end{align} 

If $A_1$'s $i$-th qubit does indeed have this property, 
and she chooses to test Bob's honesty on the relevant batch,
the probability of the $i$-th qubit passing the test is 
at most $1 - \delta^2$.  To see this, 
note that if (\ref{zer}) holds, the probability of passing the
test is maximized if the $i$-th state is 
\begin{align} 
( p_{\rm max} + \delta)^{\frac{1}{2}} \ket{00} + ( 1 - p_{\rm max} - \delta )^{\frac{1}{2}} \ket{11} \, ,
\end{align}
and $B_1$ declares that the $i$-th state is $\ket{\psi_0}$. 
The probability is then 
\begin{align}
\left(( p_{\rm max} ( p_{\rm max} + \delta ))^{\frac{1}{2}} + 
( ( 1-p_{\rm max}) (1-p_{\rm max} -\delta ))^{\frac{1}{2}}\right)^2 \leq 
1-\delta^2\, . 
\end{align}

However, the probability of $A_1$'s measurement outcomes is
independent of $B_2$'s actions.  
Hence this bound applies whether or not $B_2$ actually
implements a cheating strategy on the relevant batch.
Thus there must be a probability
of at least $\delta' \delta^2$ of at least one member
of the batch failing $A_1$'s tests.  
Hence, for any given $\delta, \delta' > 0$, the probability
that one of the $\approx 2^M$ batches for which $z=1$ fails $A_1$'s tests 
can be made arbitrarily close to $1$ by taking $M$ sufficiently
large.   

\subsection{Protocol VBCT3}
The protocol VBCT2 can be improved by using bit commitment subprotocols
to keep Bob's choice of state secret until he is able to compare the
values of $z$ announced by $A_1$ and $A_2$.  This eliminates the cheat-evident
attack discussed in the last section, and defines a protocol which
we conjecture is unconditionally secure.  We use the relativistic bit 
commitment protocol RBC2, defined and reviewed in Ref.\ \cite{Kent_relBC}. 

\begin{enumerate}
\item $B_1$ creates $N$ states, each being either 
  $\ket{\psi_0}=\alpha_0\ket{00}+\beta_0\ket{11}$ or
  $\ket{\psi_1}=\alpha_1\ket{00}+\beta_1\ket{11}$, with
  $\{\alpha_0,\alpha_1,\beta_0,\beta_1\}\in\mathbb{R}^+$, and
  $\alpha_i^2+\beta_i^2=1$.  The states are chosen with probability
  half each.  $B_1$ and $B_2$ both know the identity of the $i$-th state,
  since $B_1$ uses the shared random string ${\bf y}$ to make his random
  choices.  $B_1$ sends the second qubit of each state to $A_1$.  The values of
  $\alpha_0,\beta_0,\alpha_1$ and $\beta_1$ are known to both Alice
  and Bob. 
\item Alice decides whether to test Bob's honesty, which she codes by
  choosing the bit value $z=1$, or to trust him, coded by $z=0$.  She
  selects $z=0$ with probability $2^{-M}$. $A_1$ and $A_2$
  simultaneously inform $B_1$ and $B_2$ of the choice of $z$.
\item $B_1$ and $B_2$ broadcast the value of $z$ they received to one another.
\item If $B_1$ received $z=1$ from $A_1$, he sends the first qubit of
  each state to $A_1$, along with a classical bit identifying the
  state as $\ket{\psi_0}$ or $\ket{\psi_1}$.  If $B_2$ received $z=1$
  from $A_2$, he sends $A_2$ a classical bit identifying the state as
  $\ket{\psi_0}$ or $\ket{\psi_1}$.  These communications are sent
  quickly enough that Alice is guaranteed that each of the $B_i$ sent
  their transmission before knowing the value of $z$ sent to the
  other.  $A_2$ broadcasts the classical data to $A_1$ who tests that
  the quantum states are those claimed in the classical communications
  by carrying out the appropriate projective measurements.  If not,
  she aborts.  If so, the protocol restarts at step 1: $B_1$ creates a
  new set of $N$ states and proceeds as above.
\item If $z=0$, $A_2$ waits for time $\frac{D}{2}$ in the stationary
  reference frame of $B_2$ before starting a series of relativistic
  bit commitment subprotocols of type RBC2 by sending the appropriate
  communication (a list of suitably chosen random integers) to $B_2$.
  $B_2$ verifies the delay interval was indeed $\frac{D}{2}$, to
  within some tolerance.
\item $B_2$ continues the RBC2 subprotocols by sending $A_2$
  communications which commit Bob to the value of $i$ that defines the
  state to use for the coin toss.  
\item $B_1$ and $B_2$ then wait a further time $\frac{D}{2}$, by which
  point they have received the signals sent in step 3.   They then
  check that the $z$ values they received from the $A_i$ are the
  same. If not, they abort the protocol. 
\item $B_1$ and $B_2$ send communications to $A_1$ and $A_2$ which
  unveil the value of $i$ to which they were committed, and hence
  reveal the state chosen for the coin toss.  If the unveiling is
  invalid, Alice aborts.
\item $A_1$ and $B_1$ measure their halves of the $i$-th state in the
  $\{\ket{0},\ket{1}\}$ basis to define the outcome of the coin toss.
\renewcommand{\labelenumi}{(\arabic{enumi}.}
\item \label{Bchecktwo} As an optional supplementary post coin toss security test, $B_1$ asks
$A_1$ to return her qubits from all states other than the $i$-th. 
He then tests that the returned states are those originally sent, by
  carrying out appropriate projective measurements.  If the tests fail, he aborts the protocol.) 
\end{enumerate}

\subsubsection{Security against Alice}

In this modification of protocol VBCT2, there is no longer any advantage to 
Alice in cheating by arranging that one of the $A_i$ sends $z=0$ and
the other $z=1$.   Such an attack will be detected with certainty, as is
the case with protocol VBCT2.  Moreover, since Bob's chosen value of $i$ is 
encrypted by a bit commitment, which is only unveiled once the $B_i$ 
have checked that the values of $z$ they received are identical, 
Alice gains no information about Bob's chosen bias from the attack.  
The bit commitment subprotocol RBC2 is unconditionally
secure against Alice \cite{Kent_relBC}, since the communications she
receive are, from her perspective, uniformly distributed random strings.   

(As in the case of VBCT2, technically speaking, Alice has another possible attack:
she can follow the protocol honestly up to step \ref{Bchecktwo} and then,
once she learns Bob's chosen state, intentionally try to fail
Bob's tests by altering her halves of the remaining states in some
way.  By so doing, she can cause the protocol
to abort after the coin toss outcome is known.  
Again, though, this gives her no advantage.) 

The protocol therefore presents Alice with no useful cheating attack.

\subsubsection{Security against Bob}

Intuitively, one might expect the proof that VBCT2 is secure against Bob
to carry over to a proof that VBCT3 is similarly secure, for the following reasons.
First, the only difference between the two protocols 
is that Bob makes a commitment to the value of $i$ rather than announcing
it immediately,
Second, when the bit commitment protocol RBC2 is used, as here, just for 
a single round of communications, it is provably unconditionally secure 
against general (classical or quantum) attacks by Bob.  

To make this argument rigorous, one would need to show that RBC2
and the other elements of VBCT3 are securely {\it composable} in
an appropriate sense: i.e.\ that Bob has no collective quantum
attack which allows him to generate and manipulate collectively data
used in the various steps of VBCT3 in such a way as to cheat. 
We conjecture that this is indeed the case, but have no proof.

\subsection{Protocol VBCT4}

Classical communications and information processing are 
generally less costly than their quantum counterparts, 
so much so that in some circumstances it is reasonable
to treat classical resources as essentially cost-free
compared to quantum resources.   
It is thus interesting to note the existence of
a classical relativistic protocol for VBCT, 
which is unconditionally secure against classical attacks, and 
which we conjecture is unconditionally secure
against quantum attacks.   The
protocol requires Alice and Bob each to have 
two appropriately located agents, $A_1$, $A_2$ and $B_1$, $B_2$.  

\begin{enumerate}
\item Bob generates a $2M\times N$ matrix of bits such that each row
  contains either $\alpha_0^2 N$ zero entries or $\alpha_1^2 N$ zero
entries, these being positioned randomly throughout the row.
The rows are arranged in pairs, so that, for $m$ from $0$ to $(M-1)$,
either the $2m$-th row contains $\alpha_0^2 N$ entries 
and the $(2m+1)$-th contains $\alpha_1^2 N$, or vice versa. 
This choice is made randomly, equiprobably, and independently for each pair.
The matrix is known to both $B_1$ and $B_2$ but kept secret from Alice. 
\item Bob then commits each element of the matrix separately to Alice
  using the classically secure relativistic bit commitment subprotocol RBC2 
  \cite{Kent_relBC}, initiated by communications between $A_2$ and $B_2$. 
\item $A_1$ then picks $M-1$ pairs at random.  She asks $B_1$ to
  unveil Bob's commitment for all of the bits in these pairs of rows.  
\item The RBC2 commitments for the remaining bits are sustained while
  $A_1$ and $A_2$ communicate to verify that each unveiling
  corresponds to a valid commitment to either 0 or 1.  Alice also
  checks that each unveiled pair contains one row with $\alpha_0^2 N$
  zeros and one with $\alpha_1^2 N$ zeros.  If Bob fails either set
  of tests, Alice aborts.
\item If Bob passes all of Alice's tests. $B_1$ picks the remaining
  row corresponding to the bias he desires, and $A_2$ simultaneously picks a
  random column.  They inform $A_1$ and $B_2$ respectively, thus identifying
  a single matrix element belonging to the intersection.
\item Bob then unveils this bit, which is used as the outcome of the
  coin toss.  The remaining commitments are never unveiled.
\end{enumerate}

\subsubsection{Security} 

The above protocol shows that, classically, 
bit commitment can be used as a subprotocol to achieve VBCT.  
The proof that RBC2 is unconditionally 
secure against classical attacks \cite{Kent_relBC}
can be extended to show that Protocol VBCT4 is similarly secure.
RBC2 is conjectured, but not proven, to be secure against general
quantum attacks.  We conjecture, but have no proof, that 
the same is true of Protocol VBCT4.

\section{Summary}

We have defined the task of variable bias coin tossing (VBCT), illustrated
its use with a couple of applications, and presented 
four VBCT protocols.  Of these the first, VBCT1, allows VBCT for a limited range of
biases, and is unconditionally secure against general quantum attacks. 
The second protocol, VBCT2, is defined for any range of biases and guarantees 
cheat-evident security against general quantum attacks.  The third, VBCT3, extends the second
by using a relativistic bit commitment subprotocol, and 
we conjecture that it is unconditionally secure against
general quantum attacks.  

The fourth protocol, VBCT4, is classical, and is based on multiple
uses of a classical relativistic bit commitment scheme which
is proven secure against classical attacks.  It can be shown
to be unconditionally secure against classical attacks.   
The relevant relativistic bit commitment scheme is conjectured secure
against quantum attacks, and we conjecture that this is also 
true of Protocol VBCT4.   

Variable bias coin tossing is a simple example of a random 
one-input two-sided secure computation.   
The most general such computation is what we have termed a 
variable bias $n$-faced die roll.  In this case, there is a finite range of 
$n$ outputs, with each of Bob's inputs leading to a different probability
distribution over these outputs.  In other words, Bob is effectively
allowed to choose one of a fixed set of biased $n$-faced dice to
generate the output, while Alice is guaranteed that Bob's chosen dice
is restricted to the agreed set.

The protocols VBCT2, VBCT3 and VBCT4 can easily be generalized to
protocols defining variable bias $n$-faced die rolls.
Thus, to adapt protocols VBCT2 and VBCT3 to variable bias die rolling, 
we require Bob to choose a series of states 
from the set $\{\ket{\psi_i}= \sum_{j=0}^{n-1} \alpha^j_i \ket{jj}
\}_{i=1}^r$, where $r$ is the number of dice in the allowed set and 
where $(\alpha^j_i)^2$ defines the probability of outcome $j$
for the $i$-th dice (we take $\{\alpha^j_i\}$ to be real and
positive).  The protocols then proceed similarly to those given above,
defining protocols which we conjecture to be cheat-evidently secure
and unconditionally secure respectively. 

To adapt protocol VBCT4, we require that the matrix rows contain appropriate
proportions of entries corresponding to the various possible die roll
outcomes.  We conjecture that this protocol is unconditionally secure. 

As we noted earlier, variable bias $n$-sided die rolling is the most general
one-input random two-sided two party single function computation.
Our conjectures, if proven, would thus imply that all such 
computations can be implemented with unconditional security.   

\acknowledgments
RC gratefully acknowledges an EPSRC research studentship
and a research scholarship from Trinity College, Cambridge. 
This work was supported by the project PROSECCO (IST-2001-39227) of
the IST-FET programme of the EC.

\end{document}